\documentclass{article}
\usepackage{epsfig}
\usepackage[centertags]{amsmath}
\usepackage{graphicx}
\usepackage{url}

\begin{document}
\title{Coherent Control of Trapped Bosons}
\author{Analabha Roy and L.E. Reichl \\
Center for Complex Quantum Systems\\
and\\
Department of Physics\\
The University of Texas at Austin, Austin, Texas 78712\\}
\date{\today }
\maketitle

\begin{abstract}
We investigate the quantum behavior of a mesoscopic two-boson system produced by number-squeezing ultracold gases of alkali metal atoms. The quantum Poincare maps of the wavefunctions are affected by chaos in those regions of the phase space where the classical dynamics produces features that are comparable to $\hbar$. We also investigate the possibility for quantum control in the dynamics of excitations in these systems. Controlled excitations are mediated by pulsed signals that  cause Stimulated Raman Adiabatic passage (STIRAP) from the ground state to a state of higher energy. The dynamics of this transition is affected by chaos caused by the pulses in certain regions of the phase space. A transition to chaos can thus provide a method of controlling STIRAP.
\end{abstract}
\section{\label{sec:1} Introduction}

There have been significant advancements in  techniques for cooling and trapping ultracold atom gases in recent years, facilitating the experimental realization of Bose-Einstein condensation in  dilute alkali gases (specifically $^{87}Rb$ and $^{23}Na$) in 1995~\cite{weiman}~\cite{weiman:cornell}~\cite{ketterle}~\cite{ketterle2}. Since then, numerous studies of these condensates have been accomplished. In addition, experiments have been conducted that have obtained boson systems in number squeezed states from ultracold gases of alkali atoms in optical traps~\cite{raizen}. Thus, it is now possible to create a mesoscopic two-boson system of Sodium or Rubidium. Individual alkali atoms from a BEC reservoir can be subjected to quantum tweezers (Gaussian lasers exploiting the Landau-Zener tunnelling between the reservoir levels and the levels in the laser beam)~\cite{diener}. Other recent methods include number-squeezing the BEC itself by "culling" atoms from a trapped condensate down to a sub-poissonian regime, making the number uncertainty small enough to be ignored~\cite{raizen}.

A two-boson system can be subjected to a micrometer-scale double well potential using coherent laser beams. The methods that are applicable in the required length scales are overlapping cross-sections of Gaussian lasers~\cite{dudarev:entanglement}. Numerous methods can be used to control the extent of the overlap, such as an Acousto-Optical Modulator (AOM) vibrating with two sound waves~\cite{ketterle:aom}. Each sound wave causes an impinging laser beam with a Gaussian cross section to Bragg-diffract at an angle. The separation between the two diffracted beams can be adjusted with relative ease, and consequently so can the length scale of the double well that is generated by focussing the two diffracting beams into parallel beams (the double well lies in the direction lateral to the propagation)~\cite{ketterle:aom}. Other applicable techniques for double well generation are small volume optical traps formed by multiple lasers propagating as Gaussian sheets~\cite{raizen}, and blue-detuned far-off-resonant laser light to add a potential hill in the middle of a cigar-shaped magnetic trap~\cite{andrews:science}.

A two-boson system, in an optical trap, can be subjected to stimulated Raman scattering. As we shall show, coherent population transfer from the ground state into one of the excited states can be achieved using radiation pulses.  Time-modulated (i.e pulsed) electric fields or laser radiation pulses, applied sequentially in a counter intuitive manner,  can be used to facilitate this process. If the time scale of the pulse modulation is sufficiently large, the Raman process is adiabatic  (called Stimulated Raman Adiabatic Passage or STIRAP)~\cite{stirap:hioe} ~\cite{stirap:review}. 

The analysis of traditional STIRAP systems involves  relatively weak radiation pulses and the rotating wave approximation on the three involved levels~\cite{stirap:review}. More recently, multilevel transitions  have been performed and  Floquet theory has been used  to tell us how the population evolves in time~\cite{na-reichl:pbox}~\cite{na-reichl:mol-rot}~\cite{holder-reichl:avoidedcross}~\cite{na-reichl:isomer}. 

We investigate the appearence of avoided crossings contributed by resonating levels other than the ones that the radiation pulses connect. These avoided crossings in the Floquet eigenphases appear due to level repulsion caused by a loss of symmetry/degeneracy (actual crossings)~\cite{reichl}, and affect the statistical properties of the spectra, bringing them close to that predicted by random matrix theory. These are connected with the dynamics of the underlying classical system, which undergo a transition from KAM tori to chaos in this region of the paramter stace~\cite{reichl}. Thus, this work also demonstrates the quantum effects of chaos, induced by the radiation, on multilevel transitions in a 2-boson system.

In the following sections, we describe the behavior of two bosons in a one-dimensional double-well potential which we shall model as detailed in Section 2. In Section 3 we will describe the classical dynamics, where the pseudopotential interaction in one-dimension is approximated by a Gaussian potential. In Section 4, we will discuss the quantum eigenstates of this system and compare the quantum phase space of the eigenstates with the classical phase space. In Section 5, we will proceed to describe the dynamics of excitations of this system which will be driven by two sinusoidal electric field pulses applied in sequence. The frequencies of the fields are chosen to connect specific undriven energy levels of the system.  Section 5.1 poses the Schr\"odinger equation for the quantum dynamics for this system. Section 5.2 introduces Floquet theory and the numerical methods implemented to evaluate the Floquet matrix. Section 6 discusses a specific set of parameters for the general STIRAP dynamics discussed in Section 5. In subsections 6.1-6.3, we look at the eigenvalue spectra of the Floquet matrices for two amplitudes and use them to explain the actual dynamics of the system obtained from \textit{ab initio} numerical solutions of Schr\"odinger's equation with the same parameters. We repeat these numerical methods for a different set of parameters in Section 7, where the wells are slightly deeper and an additional resonance exists in the eigenvalue spectrum which affect the coherent excitations. For large pulse amplitudes, the dynamics of the STIRAP process is also noticeably affected by the presence of chaos in the stronger pulse amplitudes in both cases. Concluding remarks are made on Section 8.
\section{\label{sec:2} The Basic Model}
Our system consists of  two alkali metal bosons confined to a double-well optical potential. The two bosons can be obtained from  a cold-atom system confined in a Magneto-Optical Trap (MOT) by number squeezing, followed by laser culling in an optical dipole trap (ODT). If the optical laser is far-detuned from the internal atomic resonances, they can be treated as point particles. The effective interaction between the bosons, in three dimensions, is obtained in the long wavelength approximation to be 
\begin{equation}
u^{3d}({\bf x}_1-{\bf x}_2)= \frac{4\pi\hbar^2 a_s}{m} \delta({\bf x}_1-{\bf x}_2),
\end{equation}
where ${\hbar}$ is Planck's constant, $a_s$ is the s-wave scattering length and ${\bf  x}_i = \left(  x_i,y_i,z_i \right) $ is the displacement of the $i$th particle~\cite{metcalf:vanderstraten}~\cite{pethick:bec}. Therefore, the energy eigenvalues of the two particle system are given by the Schrodinger equation 
\begin{multline}
\left[ -\frac{\hbar^2}{2m}\nabla^2_1 +V_e(x_1)-\frac{\hbar^2}{2m}\nabla^2_2 +V_e(x_2)\right] 
\Psi_{E_j}({\bf x}_1, {\bf x}_2) \\
 +  \left[V_T({\bf x}_1)+V_T({\bf x}_2)\right]\Psi_{E_j}({\bf x}_1, {\bf x}_2) \\
 +  \frac{4\pi\hbar^2 a_s}{m} \delta^3({\bf x}_1-{\bf x}_2)\Psi_{E_j}({\bf x}_1 ,{\bf x}_2) 
= E_j \Psi_{E_j}({\bf x}_1,{\bf x}_2) ,
\label{eq:3dsys}
\end{multline}
where $m$ is the mass of the particles, $V_e(x_i)$ is the optical double-well potential and depends only on a single coordinate, and $V_T({\bf x}_i)$ is the potential profile of the MOT. $\Psi_{E_j}({\bf x}_1,{\bf x}_2)={\langle}{\bf x}_1,{\bf x}_2|E_j{\rangle}$ is a symmetrized  energy eigenstate of the interacting two-particle system with energy $E_j$.

The system can be confined in two spacial (radial) directions so that the essential dynamics occurs in the $x$ - direction. This can be achieved by using anisotropic magnetic traps with high aspect ratio~\cite{olshanii:1d},  where the radial trap frequencies are considerably larger than the  axial frequency~\cite{petrov:1d}. In that case, the radial part of the wavefunction is is not affected by the interaction and the energy eigenstate can be decomposed into an axial part (along the x-axis) and a radial part such that
\begin{equation}
\Psi_{E_j}({\bf x}_1, {\bf x}_2) =\psi{_j}(x_1,x_2) \phi_\bot ({\bf x}_{1\bot},{\bf x}_{2\bot}),
\end{equation}
where ${\bf x_{i\bot}}=(y_i,z_i)$ and  $\phi_\bot ({\bf x}_{1\bot},{\bf x}_{2\bot})$ denotes the noninteracting ground state of the system in the $\lbrace y, z\rbrace$ plane.  If we assume the trap to be a radial harmonic oscillator of stiffness $\omega_s$ then, multiplying Eq.~\eqref{eq:3dsys} by $\phi^*_\bot ({\bf x}_{1\bot},{\bf x}_{2\bot})$ and integrating in all radial coordinates reduces the Hamiltonian to a form which describes motion along the axial direction,
\begin{equation}
\left[ \frac{p^2_1}{2m}+V_e(x_1) + \frac{p^2_2}{2m} + V_e(x_2)+u(x_1-x_2) \right] \psi{_j}(x_1,x_2)
=E_j \psi_{j}(x_1,x_2),
\end{equation}
where
\begin{equation}
u(x_1-x_2) = 4a_s \omega_s \hbar \delta(x_1-x_2)
\end{equation}
and the energy $E_j$ contains contributions from the radial ground state. Thus, the system is effectively one dimensional.

We will consider the case of  two identical bosons  confined to a quartic double well potential.
The total Hamiltonian for the system is 
\begin{equation}
H =\frac{p^2_1}{2m}+\frac{p^2_2}{2m}+v(-2a x_1^2+b x_1^4) +v(-2a x_2^2+b x_2^4)+4a_s \omega_s \hbar \delta(x_1-x_2) ,
\end{equation}
where $p_i$ is the momentum of the $i$th particle ($i$= 1,2), $x_i$ is the position of the $i$th particle along the x-axis,  $v$ determines the depth of the double well potential, and $a$ and $b$ determine its spatial extent.

It is useful to write the Hamiltonian in terms of dimensionless parameters $(p_i', x_i',H')$. We  introduce a unit of length, $L_u=50 nm$, which (as we will see) is appropriate to the systems we consider here.  Then let  $x'_i=\frac{x_i}{L_u}$, $a=L_u^{-2}$, $b=L_u^{-4}$, $p'_i=\frac{p_i}{p_u}$, $H'=\frac{H}{E_u}$,  $V_0=\frac{v}{E_u}$, $U_0=\frac{4a_s \omega_s \hbar}{E_u}$ and $t'=\frac{t}{T_u}$ where $P_u=\frac{\hbar}{L_u}$, $E_u=\frac{\hbar^2}{2mL^2_u}$ and $T_u=\frac{2mL^2_u}{\hbar}$. If we now drop the primes on the dimensionless parameters, we obtain
\begin{equation}
H =p^2_1+p^2_2+V_0 (-2 x_1^2+ x_1^4) +V_0 (-2 x_2^2+ x_2^4)+U_0 \delta(x_1-x_2) .
\label{eq:hamscale}
\end{equation}
The unit of length, $L_u = 50 nm$, is two orders of magnitude below typical length scales in double wells created with  Gaussian lasers~\cite{dudarev:entanglement}, as well as Gaussian sheet lasers, or  lasers diffracted off of acousto-optical modulators vibrating with two sound waves~\cite{ketterle:aom}. Nonetheless, this length scale should be attainable in these setups. For instances, well dimensions in a double well generated using an AOM are proportional to the difference between the acoustic frequencies (typically $10$ MHz ~\cite{ketterle:aom}), producing a barrier length in the micrometer range. It should be relatively straightforward to reduce that frequency difference by two orders of magnitude and adjusted it to produce a double well of a barrier length of $100 nm$, or an $L_u$ of $50 nm$, as is required here. The lifetime of the magnetic traps in these experiments is approximately $3-4$ s. If we use $^{87}$Rb alkali atoms as our bosons then the value of $T_u$ is about $7 \mu s$, making typical trap lifetimes (about $5 s$) translate to $800,000$ units of $T_u$. The characteristic energy scale $E_u$ is very small, and corresponds to photons with frequency of about $24 KHz$. Therefore,  monochromatic coherent radiation at around this frequency is necessary for STIRAP excitations, and can be obtained using noble gas masers~\cite{maser}.   Fig.~\ref{fig:doublewell} shows a plot of the quartic double well $V(x)=V_0 (-2 x^2+ x^4)$ for  well depth $V_0=4.91345043$.

\section{\label{sec:3} Classical Dynamics of the Undriven System}

In order to study the classical dynamics of the system governed by the Hamiltonian in 
Eqn.~\eqref{eq:hamscale} the contact potential $U(x_1-x_2)=U_0 \delta(x_1-x_2)$ can be replaced by   a Gaussian shaped potential of suitably chosen width $\sigma_c$ such that 
\begin{equation}
U_G(x_1-x_2)=\frac{U_0}{\sigma_c \sqrt{2\pi}} e^{-\frac{(x_1-x_2)^2}{2\sigma^2_c}}.
\end{equation}
We have noticed no discernable difference in the quantum case between using the Gaussian for the interaction and using the delta function provided that the width of the  Gaussian $\sigma_c$ is sufficiently small. Too small a width generates unresolvable errors in the numerical solution to the classical dynamics and so an optimum width was chosen at $\sigma_c = 0.005$. 

In the absence of a time dependent external field, the system is conservative, has two degrees of freedom, and energy  $E_0$ is constant. Therefore,  the system is confined to a three dimensional surface in a four dimensional phase space. Fig.~\ref{fig:classicalpncr} shows Poincare surfaces of section of the classical phase space of this system for $V_0 = 4.91345043$ and $U_0 = -1.0 $ (attractive interaction) and $E_0$ fixed. The momentum $p_1$ and displacement $x_1$ of particle-1 are plotted each time the trajectory of particle-2 crosses the point  $x_2 = 1$  with $p_2 > 0$ such that $E_0$ is fixed and therefore they show the behavior of  momentum and position of particle-1 for that energy. Energy conservation places bounds on the values of  the coordinates $p_1, p_2, x_1, x_2$ so the trajectories are confined to a finite region of the phase space.

\subsection{Relevant Energies of the Quantum System}

Figs.~\ref{fig:classicalpncr}.a through~\ref{fig:classicalpncr}.c are surfaces of section for three different energies $E_1 = -3.7195$, $E_2 = -2.66655$ and $E_4 = 2.5986$, respectively. The energies were chosen to match quantum energy levels that will be connected by STIRAP pulses. The numerical integration was done by the $4^{th}$ order implicit Runge Kutta (Prince Dormand) method~\cite{rkutta:pd} using the appropriate subroutine in the GNU Scientific Library~\cite{galassi:gsl}. The integration was done non-adaptively, with a fixed temporal stepsize of $10^{-3}$.
In Fig.~\ref{fig:classicalpncr}.a ($E_0 = -3.7195$), we can see several regions of interest. There are three  kinds of dynamics at this energy vis-a-vis the relative energies of each particle. In the case that the particles are in separate wells, trajectories exist where they don't see each other and are therefore the same as that of a single particle in a double well. They are seen as large KAM (Kolmogorov-Arnold-Moser) tori around $x_1 = -1$ in Fig.~\ref{fig:classicalpncr}.a.1. These trajectories lie between two chaotic regions. Both chaotic regions are produced by the case when the energy of one particle is set to a positive value (thus taking it above the wells), and the energy of the other particle is decreased so that they both add up to $E_0$ (the particles being kept sufficiently far apart at $t=0$ so as to make the interaction negligible at that time). The resultant dynamics is chaotic due to the interaction experienced by the particles when they approach each other during motion. The cases when the particle being strobed has high momentum cause the chaos at the separatrix coupling both wells.
The island seen immediately around $x_1 = -1$ in Fig.~\ref{fig:classicalpncr}.a.1 is the result of a bifurcation that occurs at lower energies, and will be discussed in the next section.

The case when both particles are in the same well are seen as the highly elongated tori around $x_1 = 1$ in Fig.~\ref{fig:classicalpncr}.a.2 due to the interaction between the particles. In Fig.~\ref{fig:classicalpncr}.b.1, we note the disappearance of the bifurcated island immediately around $x_1 = -1$, the result of a bifurcation in reverse (as energy is increased).  The left-right asymmetry observed at lower energies is  reduced at the energy increases and (apart from the elongated tori), disappears in Figs.~\ref{fig:classicalpncr}.c.  The region of chaos seen deep  inside  the potential  wells of Fig.~\ref{fig:classicalpncr}.a.1 merges with the region of chaos at the separatrix for higher the value of energy in Fig.~\ref{fig:classicalpncr}.c.1. 

\subsection{Lower Energies}

Figs.~\ref{fig:bifurcation}.a through~\ref{fig:bifurcation}.f are surfaces of section for six energies,  all less than the quantum ground state energy, and shown in decreasing order in energy.  Only the $x_1<0$ half of the phase space is shown. For sufficiently low energies, the two particles either oscillate in two wells independently or together in the same well. The former case is seen in Fig.~\ref{fig:bifurcation}.f where the periodic motion of one particle about $x_1 = -1$ is visible at energy $E = -5$. As we increase the total energy, it becomes possible for one particle to break it's confinement for certain values of it's initial coordinates and influence the dynamics of the other particle through the interaction. This produces prominent chaotic behavior as seen in the figures. Increasing the total energy further from $E = -4.5$ to $E = -4.0$ (Figs.~\ref{fig:bifurcation}.d to~\ref{fig:bifurcation}.c) produces a bifurcation as the total energy is increased further (see Figs.~\ref{fig:bifurcation}.c to ~\ref{fig:bifurcation}.a).

\section{\label{sec:4} Quantum Mechanics of the Interacting System}

In this section, we discuss the quantum mechanics of two interacting bosons whose Hamiltonian is given by Eq.~\eqref{eq:hamscale}.  We first discuss the basis used to diagonalize the Hamiltonian. We then show configuration space and phase space plots of the key eigenstates of the system. 

\subsection{Diagonalization of the Hamiltonian}

In order to diagonalize the Hamiltonian in Eq.~\eqref{eq:hamscale}, we use the eigenstates of two noninteracting bosons in a hard-wall box as a representation to formulate the matrix elements.  We choose a box of width  $L=3.5$ (in dimensionless units)   so that an adequate balance is achieved between truncability and accuracy. The Hamiltonian  for a single particle in a box is $h=p^2$  $\forall$ $x\leq|L|$ and the energy eigenstates are given by 
\begin{equation}
\phi_n(x)=\langle x|n,x\rangle=\frac{1}{\sqrt{L}} \sin{\biggl[}{\frac{n\pi}{2}(\frac{x}{L}-1){\biggr]}},
\label{eq:pboxfn}
\end{equation}
where $n=1,2,...\infty$. 

The 2-particle boson states of the box system are obtained by symmetrizing the 2-particle states to obtain a complete orthonormal basis of symmetrized 2-boson states:
\begin{equation}
{\langle}x_1,x_2\vert n_1,n_2{\rangle} ^{(s)}=\frac{1}{\sqrt{2(1+\delta_{n_1,n_2})}} 
[{\langle}x_1|n_1\rangle{\langle}x_2|n_2\rangle +{\langle}x_1|n_2\rangle{\langle}x_2|n_1\rangle ].
\label{eq:symm}
\end{equation}
These states are then used to create a Hamiltonian matrix from Eq.~\eqref{eq:hamscale}. The eigenvalues $E_i$ and eigenvectors $\vert E_i \rangle$ of the Hamiltonian matrix were determined numerically using the appropriate subroutine for diagonalizing real symmetric matrices in the GNU Scientific Library~\cite{galassi:gsl}.  

\subsection{Energy Eigenstates}

Fig.~\ref{fig:doublewell} shows a plot of the double well system $V(x)=V_0 (-2 x^2+ x^4)$ with well depth $V_0=4.91345043$. It also shows energy levels of the interacting system with  the contact  interaction  chosen to be attractive (so the amplitude $U_0$ is negative). This can be achieved by tuning a homogeneous magnetic field to the Feshbach Resonance of the alkali metal atoms~\cite{feshbach:resonance}~\cite{pethick:bec}.

Figs.~\ref{fig:wavefunctions}.a through  \ref{fig:wavefunctions}.c are plots of the ground state, $|E_1\rangle$, first excited state, $|E_2\rangle$,  and third excited state $|E_4\rangle$ of the interacting system, respectively. Each figure shows a plot of the probability density $|\langle x_1,x_2|E_j\rangle|^2$, as well as cross sections of the wavefunctions $\langle x_1|E_j\rangle$ for $x_2=[-1,0,1]$. The bosonic character of the states is evident from the fact that they are symmetric under $x_1 \leftrightarrow x_2$ exchange. 

We can also compute the phase space distribution of the energy eigenstates and compare this with the classical surfaces of section in Fig.~\ref{fig:classicalpncr} .  A phase space distribution of quantum states was first constructed by Wigner~\cite{wigner}.  A smoothed version of the Wigner distribution was introduced by Husimi~\cite{husimi} and has proved particularly useful for comparison of classical and quantum phase space distributions. 
In the $x-p$ representation, the Husimi Function $F_h(\bar{x_1},\bar{x_2},\bar{p_1},\bar{p_2})$for a symmetrized 2-particle  wavefunction $\psi_{E_j}(x_1,x_2)$ is defined as
\begin{multline}
F_h(\bar{x_1},\bar{x_2},\bar{p_1},\bar{p_2})=\frac{1}{\sigma_1 \sigma_2 \pi} \int \frac{dx_1}{\sqrt{2\pi}} \frac{dx_2}{\sqrt{2\pi}} \Psi_{E_j}(x_1,x_2) \\
e^{-\frac{(x_1-\bar{x_1})^2}{2\sigma^2_1}} e^{-\frac{(x_2-\bar{x_2})^2}{2\sigma^2_2}} e^{i (\bar{p_1}x_1 + \bar{p_2}x_2)},
\label{eq:husimi}
\end{multline}
where $(\bar{x_1},\bar{p_1})$ and $(\bar{x_2},\bar{p_2})$ are the centroids  of the Gaussian wave packets in the phase space.  

In order to calculate the standard deviations ${\sigma}_1$ and $\sigma_2$ , we follow the same basic prescription as is normally followed for one-dimensional Husimi functions~\cite{novaes}, where the Gaussians appearing in the Husimi function are  interpreted as harmonic oscillator coherent states. Therefore the standard deviation is the same as that of the harmonic oscillator, with the modification that the "stiffness" $\omega_d$ be $\frac{2\pi}{T}$ where $T$ is the double well period of motion, making $\sigma_i=\sqrt{\frac{2}{\omega_d V_0}}$ ($i$ is $1$ or $2$). This is generalized to two particles by choosing the single-particle energies that, when added up, come closest to the 2-particle energy.
We then proceed to calculate $\omega_d$ for each particle as a single noninteracting particle with the chosen energy level. A straightforward integration of the classical double well problem shows~\cite{reichl-appendix}
\begin{equation}
\omega_d =
\begin{cases}
\frac{2}{\sqrt{V_0}}\frac{f\pi}{K(\kappa)}& ~~\text{If $E_j<0$} \\
\frac{\pi h}{\sqrt{V_0}\kappa'K(\kappa')}& ~~\text{If $E_j\geq 0$},
\end{cases}
\end{equation}
where  $K(\kappa)$  is the complete elliptic integral of the first kind, $\kappa$ is given by
\begin{equation}
\kappa^2=\frac{2\sqrt{1+\frac{E_j}{V_0}}}{1+\sqrt{1+\frac{E_j}{V_0}}},
\end{equation}
and ${\kappa}^{'2}=1-{\kappa}^2$. 

Now that the standard deviations can be calculated, we have the prescription for numerically evaluating the full Husimi function. We cannot sketch the full four- dimensional function realistically, of course. However, we can sketch a "quantum Poincare map" of the Husimi function by strobing a particular value of $\bar{x_2}$ given $\bar{p_2} > 0$ with  the energy  classically conserved at the quantum eigenvalue. Thus, we can plot
\begin{equation}
f_h(x_1,p_1) = F_h (x_1, x_2 = +1, p_1, 
 p_2=p_2(x_1,p_1,E_j)),
\label{eq:husimi_section}
\end{equation}
where $p_2$ is determined from the condition at the unperturbed Hamiltonian  $H= E_j$ for a particular eigenstate of energy $E_j$.

Figs.~\ref{fig:husimis}.a through~\ref{fig:husimis}.c are Husimi plots of the double well system for states $|E_1\rangle$, $|E_2\rangle$ and $|E_4\rangle$. They can be compared with the corresponding classical Poincare sections in Figs.~\ref{fig:classicalpncr}.a through~\ref{fig:classicalpncr}.c. The Husimi plot of $|E_1\rangle$ is provided on Figs.~\ref{fig:husimis}.a.1 and~\ref{fig:husimis}a.2. We notice that the highest probabilities are located in the separatrix region, where there is significant chaos in the classical map (Figs. ~\ref{fig:classicalpncr}.a). However, there is a significant probability for the system to tunnel from the separatrix region to the interior near the well minima. This is where the bifurcated trajectories occur in Figs. ~\ref{fig:classicalpncr}.a.1 and~\ref{fig:classicalpncr}.a.2. All the Husimis are symmetrical, apart from the interaction resonance in each case, under phase space inversion ($x_1{\rightarrow}-x_1,~p_1{\rightarrow}-p_1$). 
The tunnelling probability from the separatrix into the wells is considerably reduced in the Husimi plot of $|E_4\rangle$ (shown in Figs.~\ref{fig:husimis}.c.1 and~\ref{fig:husimis}.c.2). 

\section{\label{sec:5} Quantum Dynamics of the Driven System}
In order to control transitions between energy states of the two boson system, we drive the 
system with  two sequential pulses of maser radiation with carrier frequency, ${\omega}_f$ (${\omega}_s$ ) for the first  (second)  pulse. These frequencies are determined by the transitions of interest. The masers are projected in the dimension of particle confinement, thus the spacial dependence of the electric field
\begin{equation}
E_i(x,t) = E_{0i} e^{j(k_ix-\omega_it)} + h.c
\end{equation}
($i=f,s$ and $j=\sqrt{-1}$) in that direction can be treated as linear, given that the wavelength of the radiation pulse(s) are in the microwave range, and the trapping length scales are $\sim 100 nm$ in the double well. Ignoring purely temporal terms that only contribute an overall phase, the interaction Hamiltonian $- D\bullet E(x,t)$ simplifies to $jE_{0i}Dk_i x e^{j\omega_i t} + h.c$, where $D$ is the atomic dipole moment. Thus, the Hamiltonian of the driven system can be written as
\begin{equation}
H(t)=H+[\epsilon_f(t)\sin(\omega_ft)+\epsilon_s(t)\sin(\omega_st)](x_1+x_2),
\label{eq:drivham}
\end{equation}
where $H$ is the Hamiltonian of the non-driven system in Eqn.~\eqref{eq:hamscale} and the amplitudes $\epsilon_i(t)$ ($i=f,s$) of the driving fields have Gaussian shape. 
\begin{equation}
\epsilon_{i}(t)= A_i e^{{\beta}(t-t_{i})^2}~~~{\rm for}~~~i=f,~s,
\label{eq:drivensystem}
\end{equation}
where $A_{f}$ ($A_s$) is the maximum amplitude of the first (second) pulse, and the dipole moment $D$ and $k_i$s have been absorbed into the $A_i$s. The duration of each pulse is controlled by the parameter $\beta=\frac{1}{2{\tau}^2}$, where ${\tau}$ is a measure of the width of each pulse (similar to standard deviation of the Gaussian). The time at which the maximum of the $i$th  pulse occurs is $t_i$. 

The  Schr\"{o}dinger equation for the 2-boson system in the presence of the driving field is
\begin{equation}
i\frac{\partial}{\partial t}\vert \psi(t)\rangle=H(t)\vert \psi(t)\rangle .
\end{equation}
Given our numerical expressions for the energy eigenstates $|E_j{\rangle}$ of the non-driven  system, we can expand $|\psi(t)\rangle$ in terms of these states so that
\begin{equation}
|\psi(t)\rangle = \sum_jc_j(t) |E_j\rangle.
\end{equation}
The Schr\"odinger equation can then be written in the form
\begin{equation}
\frac{dc_j}{dt} =-iE_jc_j(t)-i[ \epsilon_f(t) \sin(\omega_ft) 
+\epsilon_s(t) \sin(\omega_st)] \sum_{j'}d_{j,j'}c_{j'}(t),
\label{eq:schroedinger}
\end{equation}
where $c_j(t)=\langle E_j|\psi(t)\rangle$ is the probability amplitude to find the system in state $|E_j\rangle$ at time $t$ and $d_{j,j'}={\langle}E_j|(x_1+x_2)|E_{j'}{\rangle}$ denotes dipole matrix elements taken with respect to the exact energy eigenstates of the undriven system. Values of the dipole matrix elements are given in Table~\ref{tabA}.

\begin{table}
\begin{center}
\begin{tabular}{|c|ccccc}
$D^s_{ij}$ & 1& 2& 3& 4& ...\\
\hline
1& 0& -0.108& 0& 0& ...\\
2& -0.108& 0& -0.053& -0.008&  ...\\
3& 0& -0.053& 0& 0&... \\
4& 0& -0.008& 0& 0&   ... \\
5& 0& 0& 0.015& 0.002&   ... \\
6& 0& 0& 0& 0&   ... \\
7& 0.017& 0& 0& 0.015&  ... \\
8& 0& 0.003& 0& 0&   ...\\
9& 0& 0& 0& 0.001& ... \\
10& 0& 0& 0& 0&   ... \\
$\vdots$& $\vdots$& $\vdots$& $\vdots$& $\vdots$& $\ddots$\\
\end{tabular}
\caption{Dipole Matrix elements for $V_0=4.91345043$ and $U_0=-1.0$. The first $4 X 10$ values are shown here. }
\label{tabA}
\end{center}
\end{table}
\subsection{Floquet States}
For the case when the amplitude of the Gaussian pulses changes very slowly relative to the period of the carrier frequencies of the pulses, it is possible to use Floquet theory to study the dynamics of the driven system. As was shown in~\cite{na-reichl:pbox},~\cite{na-reichl:mol-rot},~\cite{na-reichl:isomer}, one can divide the time over which the pulses act into a sequence of time intervals.  During each time interval, the amplitude of the pulses is essentially constant while the carrier waves undergo many oscillations.  Consider the time window centered at time $t=t_{fix}$.  The Hamiltonian describing the dynamics during this time can be written 
\begin{equation}
H(t;t_{fix})=H+\left[\epsilon_f(t_{fix})\sin(\omega_ft)+\epsilon_s(t_{fix})\sin(\omega_st)\right](x_1+x_2).
\end{equation}
If the two frequencies ${\omega}_f$ and ${\omega}_s$  are commensurate so that $\frac{\omega_f}{\omega_s}=\frac{n_f}{n_s}$ where $n_f$ and $n_s$ are integers, then the Hamiltonian $H(t;t_{fix})$  is time-periodic with a period 
\begin{equation}
T=\pi(\frac{n_f}{\omega_f}+\frac{n_s}{\omega_s}).
\label{eq:effperiod}
\end{equation}
Because $H(t;t_{fix})$ is time periodic, Floquet theory can be used to  analyze the dynamics of the system during the time window centered at $t=t_{fix}$.

Let us assume that the Schrodinger equation, $i\frac{{\partial}}{{\partial}t}|\psi(t){\rangle}=H(t;t_{fix})|\psi(t){\rangle}$ has a solution of the form
\begin{equation}
|\psi(t)\rangle = e^{-i\Omega_{\alpha}t}|\phi_{\alpha}(t)\rangle ,
\end{equation}
where the state $|\phi_{\alpha}(t)\rangle$ is time-periodic with period $T$ and the phase $\Omega_{\alpha}$ is real.  If this is substituted into the Schrodinger equation we obtain the following eigenvalue equation
\begin{equation}
(H(t;t_{fix})-i\frac{\partial}{\partial t})|\phi_{\alpha}(t)\rangle = \Omega_{\alpha}|\phi_{\alpha}(t)\rangle.
\end{equation}
The state $|\phi_{\alpha}(t)\rangle$ is the $\alpha$th Floquet eigenstate and $\Omega_{\alpha}$ is the $\alpha$th Floquet eigenphase.  The quantity $H_F(t){\equiv}H(t;t_{fix})-i\frac{\partial}{\partial t}$ is a Hermitian operator and is called the Floquet Hamiltonian. The Floquet eigenstates $|\phi_{\alpha}(t)\rangle$  form a complete orthonormal basis and the Floquet eigenphases $\Omega_{\alpha}$ are conserved quantities~\cite{reichl}.

The  state of the boson system at time $t$ can be expanded in a Floquet spectral decomposition as follows
\begin{equation}
|\psi(t)\rangle = \sum_{\alpha}A_{\alpha}e^{-i\Omega_{\alpha}t}|\phi_{\alpha}(t)\rangle
= \sum_{\alpha}\langle \phi_{\alpha}(0)|\psi(0)\rangle e^{-i\Omega_{\alpha}t}|\phi_{\alpha}(t)\rangle .
\end{equation}
Because the Floquet eigenstates are time-periodic, the state of the system at time $t=T$ is given by
\begin{equation}
|\psi(T)\rangle = \sum_{\alpha} e^{-i\Omega_{\alpha}T}|\phi_{\alpha}(0)\rangle\langle \phi_{\alpha}(0)|\psi(0)\rangle .
\end{equation}
The Floquet evolution operator is therefore given  by $U_F(T)$ where
\begin{equation}
U_F(T) = \sum_{\alpha} e^{-i\Omega_{\alpha}T}|\phi_{\alpha}(0)\rangle\langle \phi_{\alpha}(0)|
\end{equation}
and is a unitary operator.  When the operator $U_F(T)$ acts on the state of the boson system it evolves it forward in time by one period of the driving field.

It is possible to compute the matrix elements of the Floquet evolution operator using energy eigenstates of the undriven system as the basis functions. Thus, 
\begin{equation}
U_{j,j'}(T) = \langle E_j|U_F(T)|E_{j'}\rangle 
= \sum_{\alpha} e^{-i\Omega_{\alpha}T}\langle E_j|\phi_{\alpha}(0)\rangle\langle \phi_{\alpha}(0)|E_{j'}\rangle.
\label{eq:Floquet_mat}
\end{equation}
The $\alpha$th eigenvalue of this matrix is $e^{-i\Omega_{\alpha}T}$ and the $\alpha$th eigenvector is given by a column matrix  with entries ${\langle}E_j|\phi_{\alpha}(0)\rangle$. Since we will only have numerical expressions for the eigenvalues $e^{-i\Omega_{\alpha}T}$, we can only determine the value of the \textit{eigenphases} $\Omega_{\alpha}$   modulo $2\pi$.

The numerical computation of the Floquet matrix $U_{j,j'}(T)$  is achieved as follows. Each column of the matrix can be constructed by solving the Schr\"odinger equation (with Hamiltonian $H(t;t_{fix})$)  for one period $T$.  Each column of the initial state starts with a single entry ${\langle}E_j|\phi_{\alpha}(0)\rangle=1$  for $j=\alpha$ and ${\langle}E_j|\phi_{\alpha}(0)\rangle=0$ otherwise.  The integration is done $N$ times with $\alpha$ ranging from $\alpha=1$ to $\alpha=N$.  The numerical integrations were performed using the appropriate subroutine for the $4^{th}$ order Runge-Kutta-Fehlberg method~\cite{rkutta:pd} from the GNU Scientific Library~\cite{galassi:gsl}. Each different initial condition yields one column of the Floquet matrix at time $t=T$. Performing these $N$ integrations yields an $N{\times}N$ Floquet evolution matrix. Numerically diagonalizing this matrix gives us the Floquet eigenphases and eigenstates. The numerical diagonalization of the non-Hermitian Floquet matrices were performed using the appropriate routine in the IBM{\texttrademark}  Engineering and Scientific Subroutines Library (ESSL)~\cite{ibm:essl}. This process can be performed for each value of $t_{fix}$ and the resulting eigenphases and eigenstates plotted as functions of $t_{fix}$.

In order to determine the appropriate numerical truncation for the evaluation of the Floquet matrices, we used iteratively increasing values of $N$ and checked the components of the Floquet states at the value(s) of $t_{fix}$ when  the STIRAP amplitudes were the largest ie at $t=t_f$ or $t_s$ (see Eq.~\eqref{eq:drivensystem}) until the higher components were too small to contribute to the dynamics.We chose $N=25$ as the final truncation.

In the subsequent section we show that coherent  transitions between symmetrized two-particle boson states can be achieved for this  system. Because of the sparsity on nonzero dipole matrix elements, the simplest transition, induced by the laser pulses,  is from the ground state $|E_1{\rangle}$ to the fourth level $|E_4{\rangle}$, via the intermediate state $|E_2{\rangle}$.  We show the behavior of the system for three different amplitudes of the radiation pulses. 

\section{\label{sec:6} Case 1: STIRAP Ladder, First Pulse $2\leftrightarrow4$ , Second Pulse $1\leftrightarrow2$}

Fig.~\ref{fig:doublewell} shows the energy levels of the double well system for well depth $V_0=4.91345043$ and interaction strength  $U_0=1.0$. The value of $V_0$ was chosen so the radiation pulses would have carrier wave frequencies ${\omega}_f$ and ${\omega}_s$ commensurate with each other and  so that  ${\omega}_f$ (${\omega}_s$) would be equal to the energy spacing $E_4-E_2$ ($E_2-E_1$), with a high degree of precision.  The energy levels shown in the figure are the exact energy eigenvalues of the undriven two-boson symmetrized system.  
 
We plan to use radiation pulses to induce a coherent transition of the two boson system from its ground state $|E_1{\rangle}$ to  the excited state $|E_4{\rangle}$. At $t=0$ the first pulse connects the levels $E_2$ and $E_4$ with zero detuning. The second pulse connects the  levels $E_1$ and $E_2$.  The ratio $\frac{w_f}{w_s}=\frac{5}{1}$ to eight decimal places.

The dipole moments of these transitions have very different values.  The dipole moment 
$d_{2,4}={\langle}E_2|(x_1+x_2)|E_4{\rangle}$ that couples the states $|E_2{\rangle}$ and $|E_4{\rangle}$  is two 
orders of magnitude smaller than dipole moment $d_{1,2}={\langle}E_1|(x_1+x_2)|E_2{\rangle}$ that couples the states 
$|E_1{\rangle}$ and $|E_2{\rangle}$ (See Table 1).  From Eqs.~\eqref{eq:drivham} and~\eqref{eq:drivensystem}, the amplitude 
of the first pulse is given by $A_f d_{2,4}$ and the amplitude of the second pulse is given by $A_s d_{1,2}$. 
Because the dipole coupling of the first pulse is so much smaller than that of the first pulse, we will make the electric field amplitude, $A_f$,  of  the first pulse considerably larger than that of the second pulse, $A_s$  so that 
\begin{equation}
{\epsilon}_0{\equiv}A_fd_{2,4}=A_sd_{1,2}.
\label{eq:amps}
\end{equation}
The amplitudes for the two radiation pulses are plotted in Fig.~\ref{fig:stirap}. 

The duration of each pulse can be controlled by varying the pulse width parameter  $\tau$. We let  $t_{tot}$ denote the total time over which both pulses act on the system. We choose the following values for the pulse parameters
\begin{equation}
\tau=\frac{1}{8} t_{tot},~~~t_f = \frac{1}{3} t_{tot}, ~~~{\rm and}~~~t_s = \frac{2}{3} t_{tot}.
\label{eq:pulsetimes}
\end{equation}

In the sections below, we will study the effect of these radiation pulses on the boson system  for two values of $\epsilon_0$. In both of these cases, we set a value for $\epsilon_0$ and set a suitable truncation value $N$ for the Floquet evolution matrix. 

The Floquet eigenphases lie within a {\it fundamental zone} (they are determined modulo $\omega$)   that is  taken to be $\lbrace 0,\omega\rbrace$ where $\omega=\frac{2\pi}{T}=1.05303$.  They can be plotted as a function of $t_{fix}$. For closely spaced values of $t_{fix}$, Floquet eigenstates belonging to different eigenphases, at neighboring values of $t_{fix}$,  will be orthogonal. This can be exploited to tag and follow the evolution of each eigenstate and eigenphase as a function of $t_{fix}$. 

In subsequent sections, we label each Floquet eigenphase based on it's dominant dependence on the undriven Hamiltonian eigenstates, $|E_j{\rangle}$  at $t_{fix}=0$. For the three levels, $|E_1{\rangle}$,  $|E_2{\rangle}$ and $|E_4{\rangle}$,  that are connected by the STIRAP pulses, the corresponding Floquet eigenstates have the following structure and labels:
\begin{enumerate}
\item 
The eigenphase whose corresponding Floquet eigenstate is dominated by the undriven ground state $|E_1{\rangle}$  at $t_{fix}=0$ is labelled as $\Omega_A$ and the Floquet eigenstate as $|\phi_A\rangle$.

\item
The eigenphase whose corresponding Floquet eigenstate is dominated by the undriven state $\frac{1}{\sqrt{2}}\left[|E_4\rangle - |E_2\rangle \right]$ at $t_{fix}=0$ is labelled as $\Omega_B$ and the Floquet eigenstate as $|\phi_B\rangle$.

\item
The eigenphase whose corresponding Floquet eigenstate is dominated by the undriven state $\frac{1}{\sqrt{2}}\left[|E_4\rangle + |E_2\rangle \right]$ at $t_{fix}=0$  is labelled as $\Omega_C$ and the Floquet eigenstate as $|\phi_C\rangle$.

\item
The eigenphase whose corresponding Floquet eigenstate is dominated by the undriven  state $|E_7{\rangle}$  at $t_{fix}=0$ is labelled as $\Omega_D$ and the Floquet eigenstate as $|\phi_D\rangle$.

\end{enumerate}
The symmetric and antisymmetric Floquet states  $\frac{1}{\sqrt{2}}\left[|E_4\rangle {\pm} |E_2\rangle \right]$  are induced by the first radiation pulse (which couples the states $|E_2{\rangle}$ and $|E_4{\rangle}$) even though the amplitude of the first radiation pulse may be very small. 

The results of the Floquet analysis described above can be compared to the  exact dynamics of the system obtained  by solving the full Schr\"odinger equation in Eq.~\eqref{eq:schroedinger} for the exact state of the driven system $|{\psi}(t){\rangle}$.  In solving for $|{\psi}(t){\rangle}$,  we will always start at time $t=0$ with the system in the  ground state $|{\psi}(0){\rangle}=|E_1{\rangle}$  of the undriven Hamiltonian.   We can  then  plot the probability $P_j(t)=|\langle E_j|\psi(t) \rangle|^2$ of finding the system in the undriven energy level $|E_j \rangle$ as a function of time $t$ for various values of $t_{tot}$.
We cannot show strobed Husimi plots of the Floquet states as they evolve across $t_{fix}$, nor can we show classical Poincare maps of the system during those times, since the system has five degrees of freedom during the STIRAP process.

\subsection{Pulse Amplitude $\epsilon_0 = 10$}

Fig.~\ref{fig:e_10phase}.a we plot the Floquet eigenphases as a function of $t_{fix}$  in units of the total pulse time $t_{tot}$. The eigenphases of interest are the ones involved in the STIRAP process (ie $\Omega_A$,$\Omega_B$,$\Omega_C$) which lie in the interval $\lbrace 0.488{\rightarrow} 0.5\rbrace$ in Fig.~\ref{fig:e_10phase}.a.  Fig.~\ref{fig:e_10phase}.b shows a magnification of that region. It is clear that the three levels contribute in a manner characteristic of a  traditional STIRAP ladder process  approximated by a three-level system~\cite{stirap:review}.  A three-level avoided crossing occurs at $t_{fix}\simeq 0.5t_{tot}$, and  a coherent population transfer takes place from the ground state to the third excited state. This is further confirmed by Figs.~\ref{fig:e_10states}.a and~\ref{fig:e_10states}.b. Fig.~\ref{fig:e_10states}.a  shows the evolution of  the dependence of $|\phi_A\rangle$ on the undriven energy eigenstates $|E_j{\rangle}$  as a function of time $t_{fix}$.  A population transfer occurs  from the ground state (labelled "$1$") to the fourth energy level $E_4$ (labelled ``$4$''). 

Fig.~\ref{fig:e_10states}.b  shows the actual time evolution of the state of the system $|\psi(t){\rangle}$ obtained by solving the Schr\"odinger equation as a function of $t$ for $t_{tot} = 28000$. In Fig.~\ref{fig:e_10states}.b, we plot the value of $P_j(t)=|{\langle}E_j|\psi(t){\rangle}|^2$ as a function of time $t$. The real time evolution is very close to the evolution of the Floquet eigenstate $|\phi_A{\rangle}$ as a function of $t_{fix}$. This indicates that the evolution is governed by a single Floquet eigenstate and that the process is adiabatic. The small oscillations and deviations of Fig.~\ref{fig:e_10states}.b  from Fig.~\ref{fig:e_10states}.a  can be attributed to nonadiabatic effects~\cite{berry:base}.

\subsection{Pulse Amplitudes $\epsilon_0 = 115$}

We now  set $\epsilon_0$ at a higher value of $115$.  Fig.~\ref{fig:e_115phase}.a   shows the evolution of the Floquet eigenphases as a function of $t_{fix}$. Fig.~\ref{fig:e_115phase}.b shows a magnification of the region containing eigenphases $\Omega_A$, $\Omega_B$, $\Omega_C$ and $\Omega_D$. We notice the prominence of a new Floquet state $\vert\phi_D\rangle$ (with corresponding eigenphase profile $\Omega_D$)  which,  at $t_{fix}=0$, is displaced in value from  $\Omega_A$, $\Omega_B$ and $\Omega_C$.  At $t_{fix}=0$, $\vert\phi_D\rangle$ is dominated by $\vert E_7\rangle$. This occurs due to the near-resonance between the $2-4$ transition and the $4-7$ transition (see Fig.~\ref{fig:doublewell}).

We also note, from Table~\ref{tabA}, that the $4-7$ dipole moment ($0.015$) is an order of magnitude higher than the dipole moment of the connected states $2-4$. Since the $2-4$ resonance is very close to the $4-7$ resonance, the evolution of the corresponding eigenphase $\Omega_D$ is affected and influences the evolution of the eigenphase $\Omega_A$. However no measureable avoided crossing with $\Omega_A$ occurs. Therefore,  $\Omega_A$ and $\Omega_D$ appear to cross (or under go an avoided crossing very closely spaced) and do not contribute anything significant to the dynamics. 

There are three avoided crossings that can affect transitions in the system for ${\epsilon}_0=115$.  First,  $\Omega_A$ appears  to undergo an avoided crossing with $\Omega_C$ at $t_{fix}\simeq 0.29 t_{tot}$ 
(see Fig. 10.a.1). Then the same pair of states avoid each other again at  $t_{fix}\simeq 0.35 t_{tot}$ (see Fig. 10.b). Finally, the states $\Omega_A$, $\Omega_B$, and $\Omega_C$ undergo the standard STIRAP transition at $t_{fix}\simeq 0.5 t_{tot}$.  The dependence of $\vert\phi_A\rangle$,$\vert\phi_B\rangle$, and $\vert\phi_C\rangle$ on the unperturbed energy eigenstates is shown in Fig.~\ref{fig:e_115states}.  The influence of these avoided crossings on these states is clearly seen. These avoided crossings are manifestations of classical chaos in the quantum dynamics as elaborated in the introduction.

We can now use known properties of avoided crossings to analyze this process in more detail. 
When two Floquet eigenphases $\Omega_\alpha$ and $\Omega_\beta$ approach and undergo an isolated avoided crossing, the probability $P_{\alpha \beta}$ that  the system switches from one Floquet state to the other   can be calculated from the Landau-Zener formula for two-level systems ~\cite{zener:lzformula} (note that use of this estimate for multi-level systems  assumes that other levels are not significantly involved in the avoided crossing). In our dimensionless units, the Landau-Zener probability $P_{\alpha \beta}$ is
\begin{equation}
P_{\alpha \beta}=\exp\left[-\frac{\pi ({\delta \Omega_{\alpha \beta}})^2}{2\Gamma_{\alpha \beta}}\right],
\label{eq:landauzener}
\end{equation}
where $\delta\Omega_{\alpha\beta}$ is the (minimum) spacing  between $\Omega_\alpha$ and $\Omega_\beta$ at the avoided crossing and $\Gamma_{\alpha\beta}$   is the magnitude of the  rate of change (slope) of the Floquet eigenphases in the immediate neighborhood of the avoided crossing. Thus,
\begin{equation}
\Gamma_{\alpha\beta} = {\biggl|} \frac{d\Omega_\alpha}{dt} - \frac{d\Omega_\beta}{dt}{\biggr|},
\label{eq:gamma}
\end{equation}
where  $\frac{d\Omega_\alpha}{dt}$ is the slope of the  eigenphase curve $\Omega_\alpha$ in the neighborhood of the avoided crossing. If the system switches between the two Floquet states at the avoided crossing, (if $P_{\alpha \beta}{\approx}1$),  then the energy eigenstates of the undriven system which contribute the evolution do not change significantly.  On the other hand, if $P_{\alpha \beta}{\approx}0$, then the system follows a single Floquet state through the avoided crossing, but there can be significant change in the energy eigenstates of the undriven system that contribute to the dynamics. 

The value of $\Gamma_{\alpha\beta}$ depends on the duration of the pulses $t_{tot}$ because that determines the slopes of the Floquet eigenphase curves as they enter and leave the avoided crossing. To make this explicit, we can write 
\begin{equation}
\frac{d{\Omega}}{dt}=\frac{1}{t_{tot}}\frac{d{\Omega}}{d{\tau_{ac}}},
\end{equation}
where $\tau_{ac}=t_{fix}/t_{tot}$ is the time (normalized to $t_{tot}$) at which the avoided crossing 
occurs. The quantity ${\bar \Gamma}=\frac{{d\Omega}}{d{\tau}_{ac}}$  has very weak dependence on  $t_{tot}$. We can now write the Landau-Zener transfer probability in the form
\begin{equation}
P_{\alpha \beta}=\exp\left[-t_{tot}{\gamma}_{\alpha,\beta} \right].
\label{eq:lanzen}
\end{equation}
The quantity ${\gamma}_{\alpha,\beta}=\frac{\pi ({\delta \Omega_{\alpha \beta}})^2}{2{\bar \Gamma}_{\alpha \beta}}$ has  weak dependence on $t_{tot}$. Thus, the transfer probability $P_{\alpha \beta}$ will be very small if $t_{tot}>1/{\gamma}_{\alpha,\beta}$. 

We can  compute the Landau-Zener probability for the  avoided crossings that occur for this case.  We use the information from Fig. ~\ref{fig:e_115avcrossing} to calculate ${\gamma}_{\alpha,\beta}$. For the  avoided crossing between $\Omega_C$ and $\Omega_A$, shown in Fig.~\ref{fig:e_115avcrossing}.a at $t_{fix} \sim 0.29$,  $\delta\Omega_{C,A}=0.000605$ and ${\bar \Gamma}_{C,A}=0.0929395$. Therefore,  ${\gamma}_{C,A}=6.19{\times}10^{-6}$ and we must have $t_{tot}>1.62{\times}10^5$ to have a small  probability that the system will transfer from one Floquet state to the other.  For the second avoided crossing between $\Omega_C$ and $\Omega_A$ shown in Fig.~\ref{fig:e_115avcrossing}.b  at $t_{fix} \sim 0.35$, $\delta\Omega_{C,A}=0.0032651$ and ${\bar \Gamma}_{C,A}=0.136394$. This  means that $t_{tot}>8144.8$ for an adiabatic passage (no change in Floquet eigenstate) through the avoided crossing. 

Fig.~\ref{fig:e_115timeev}.a,  which has a relatively small value of $t_{tot}$ ($t_{tot}{\sim}700$), shows no effect of these first two avoided crossings, although  it does show the effect of the three-way avoided crossing that  occurs about halfway into the total time. For this case,  the pulses appear to leave the system in a superposition of Floquet states $|{\phi}_A{\rangle}$ and $|{\phi}_C{\rangle}$.   The effect of the avoided crossing at $t_{fix}{\sim}0.29$  is also absent in Fig.~\ref{fig:e_115timeev} b, where $t_{tot}=7000$.  However, the effect of the  avoided crossing at $t_{fix} \sim 0.35$ can be seen in the Fig.  A complex mixing of states $|E_j{\rangle}$ occurs just after this avoided crossing as the large central three-way avoided crossing comes into play, and in the end the system is again left in a superposition of Floquet states $|{\phi}_A{\rangle}$ and $|{\phi}_C{\rangle}$.   The avoided crossing at $t_{fix}{\sim}0.29$ finally  starts to manifest itself in Fig.~\ref{fig:e_115timeev} (c), where $t_{tot}=72,000$ and is clearly visible in Fig.~\ref{fig:e_115timeev} d, where $t_{tot}=720,000$.  Indeed, in  Fig.~\ref{fig:e_115timeev}.d,  the system follows a single Floquet state through the entire process.  This is  confirmed by comparing the evolution of the Floquet eigenstate $|\phi_A{\rangle}$ in Fig.~\ref{fig:e_115states}.b to the exact time evolution in Fig.~\ref{fig:e_115timeev}.d.  They are essentially identical.

\section{\label{sec:7} Case 2: STIRAP Ladder, First Pulse $2\leftrightarrow4$ , Second Pulse $1\leftrightarrow2$ and nearly tuned to $4\leftrightarrow7$}

We now want to show an interesting effect that can occur in a multilevel system.  We adjust the shape of the double-well potential so that there is a  resonance between the $2 \leftrightarrow 4 $ and $4 \leftrightarrow 7$ transitions that is almost exact (to within $10^{-3}$ units of energy).
Fig.~\ref{fig:doublewell_case02} shows the energy levels of the double well system for wells that are a little deeper than in Case 1.  Here, $V_0 = 7.2912229$ and interaction strength  $U_0 = -1.0$. The energy levels shown in the figure are the exact energy eigenvalues of the undriven two-boson symmetrized system.

The classical dynamics of the system, for these deeper potential wells, is qualitatively the same as in Case 1, depicted in Figs.~\ref{fig:classicalpncr} and Figs.~\ref{fig:bifurcation}. However, deepening the wells  lowers  the quantum energies.  Nonetheless, the classical dynamics that was observed at energies $E_1$, $E_2$ and $E_4$ in Case 1 is very similar to that  seen for the corresponding  energies $E_1$, $E_2$ and $E_4$ for  Case 2.  For Case 2, that we consider in this section, the energy $E_7$ will play a significant role. In Fig.~\ref{fig:classicalpncr_case02}.a.1 ($x_1 \leq 0$) and~\ref{fig:classicalpncr_case02}.a.2 ($x_1 \geq 0$) we show surfaces of section for the classical non-driven interacting system  for an energy equal to the seventh energy level $E_7$ of Case 2.  The chaos is more spread out and the effect of a  new bifurcation can be seen.

We use radiation pulses to  induce a coherent transition of the two-boson system from its ground state in a manner similar to Case 1.  At $t=0$ the first pulse connects the levels $E_2$ and $E_4$ with zero detuning. The second pulse connects the  levels $E_1$ and $E_2$.  In this case, the ratio $\frac{w_s}{w_p}=\frac{69}{8}$. As in Case 1, there is a significant difference ($3$ orders of magnitude) between dipole moments $d_{12}$ and $d_{24}$ (see table~\ref{tabB}). Therefore, the peak amplitudes of the first and second pulses are adjusted in accordance with Eq.~\eqref{eq:amps} (see Fig.~\ref{fig:stirap}).
\begin{table}
\begin{center}
\begin{tabular}{|c|ccccc}
$D^s_{ij}$ & 1& 2& 3& 4& ...\\
\hline
1& 0&-0.103& 0& 0& ...\\
2&-0.103 & 0&0.043 &0.003 &  ...\\
3& 0&0.043 & 0& 0&... \\
4& 0&0.003 & 0& 0&   ... \\
5& 0& 0&-0.007 &-0.0004 &   ... \\
6& 0& 0& 0& 0&   ... \\
7&-0.005 & 0& 0&0.002 &  ... \\
8& 0&0.003 & 0& 0&   ...\\
9& 0& 0& 0&-0.0002 & ... \\
10& 0& 0& 0& 0&   ... \\
$\vdots$& $\vdots$& $\vdots$& $\vdots$& $\vdots$& $\ddots$\\
\end{tabular}
\caption{Dipole Matrix elements for $V_0=7.2912229$ and $U_0=-1.0$. The first $4 X 10$ values are shown here. }
\label{tabB}
\end{center}
\end{table}
In Fig.~\ref{fig:phase_case02}.a we plot the Floquet eigenphases as a function of $t_{fix}$  in units of the total pulse time $t_{tot}$. The fundamental zone  has been set to $\lbrace - \frac{\omega}{2},\frac{\omega}{2}\rbrace $, where $\omega$, the commensurate frequency, is given by $\frac{2\pi}{T}$ and $T$ is calculated from Eq.~\eqref{eq:effperiod}.  In this case, the ratio $\frac{\omega_f}{\omega_s}=\frac{69}{8}$. The eigenphases of interest are  labelled in the same manner as in case 1 ie as $\Omega_A$,$\Omega_B$,$\Omega_C$, and $\Omega_D$. These eigenphases lie in the interval $\lbrace -0.02, +0.01 \rbrace$ in Fig.~\ref{fig:phase_case02}.a.  Fig.~\ref{fig:phase_case02}.b  shows a magnification of that region. 

As we can see from Fig.~\ref{fig:phase_case02}.b,  all four eigenphases $\Omega_A$, $\Omega_B$, $\Omega_C$ and $\Omega_D$, participate in a complicated set of avoided crossings. Firstly, an avoided crossing between $\Omega_C$ and $\Omega_D$ very near $t_{fix} = 0$ causes them to switch their supports. Thus, $\vert \phi_C \rangle$ is predominantly supported by $\vert E_7\rangle$ after this crossing is avoided. The next avoided crossing of importance is the one between $\Omega_A$ and $\Omega_C$ at $t_{fix}\simeq 0.50 t_{tot}$, which causes the support of $\vert \phi_A \rangle$ to change from $\vert E_1 \rangle$ to that of  $\vert \phi_C\rangle$ viz. $\vert E_7\rangle$. Thus, a complete population transfer from  $\vert E_1\rangle$ to  $\vert E_7\rangle$ is possible. The effect of these avoided crossings and the possible behavior of the  system, as the radiations pulses pass through the system, can be seen in Fig.~\ref{fig:states_case02}. Figs.~\ref{fig:states_case02}.a through~\ref{fig:states_case02}.d show the time strobed plots of $\vert \phi_A \rangle $ through $\vert \phi_D \rangle$, respectively, analogous to Fig.~\ref{fig:e_115states} of Case 1. It is clear from Fig.~\ref{fig:states_case02}.a that this unexpected transition from energy level $E_1$ to $E_7$ should be possible to achieve, producing a marked influence of classical chaos in the quantum dynamics as elaborated in the introduction.

In order to obtain a rough estimate of the pulse time $t_{tot}$ needed to achieve true adiabatic behavior,   we apply the Landau Zener formula  Eqn.~\eqref{eq:lanzen} in the same manner as previously done in Case 1, even though it was not meant to be applicable when multiple avoided crossings are involved. For the avoided crossing of $\Omega_A$ with  $\Omega_C$ at $t_{fix}\simeq 0.50 t_{tot}$, shown in Fig.~\ref{fig:phase_case02}.b, $\delta \Omega_{A,C} = 0.0047117$ and ${\bar \Gamma}_{A,C} = 0.0403021$. Therefore $\gamma_{A,C} = 0.000865261$ and we must have $t_{tot} > 1.2 {\times} 10^3$ to have a small probability that the system transfers from $\vert \phi_A\rangle$ to $\vert \phi_C \rangle$. 

Fig.~\ref{fig:e_115timeev_case02}.a through~\ref{fig:e_115timeev_case02}.c show the actual time evolution of the system, starting from the ground state $|E_1\rangle$ at $t = 0$ for different values of $t_{tot}$. For small values of $t_{tot}$, below the threshold calculated with the Landau Zener formula (Fig.~\ref{fig:e_115timeev_case02}.a), we see a partial coherent transfer to $|E_4\rangle$, as demonstrated above. 
When $t_{tot}$ is well above threshold, as in Fig.~\ref{fig:e_115timeev_case02}.c,  complete population transfer to $|E_7\rangle$ is achieved and we appear to have reached approximately  adiabatic behavior. 
\section{Conclusions}
We have observed some unusual behavior in the classical and quantum dynamics of two bosons in a double well. Chaos in the separatrix region of the classical version of the coupled system corresponds with regions of high probability in the quantum Poincare map. However, a noticeable tunnelling has been observed from the separatrix into the individual wells. We have also demonstrated the feasibility of a controlled excitation of the system into a higher energy state using STIRAP. The STIRAP pulses destroy symmetries and produce chaos that we can detect by observing avoided crossings in the Floquet eigenphase spectrum. The chaos produced by additional resonances produce avoided crossings that can cause coherent population transfer to higher states. Thus, radiation pulses can be used to exert coherent control of the coupled boson system through chaos assisted adiabatic passages, just as has been recorded for systems with lower degrees of freedom.

\section{Acknowledgments}

The authors wish to thank the Robert A. Welch Foundation (Grant No. F-1051) for support of this work. A.R. thanks Kyunsun Na and Benjamin P. Holder for useful discussions about Floquet theory and the numerical implementation of Floquet analysis. Both authors also thank the Texas Advanced Computing Center (T.A.C.C.) at the University of Texas at Austin for the use of their high-performance distributed computing grid.

\pagebreak

\begin{figure} 
\caption{Plot of the double-well potential experienced by each boson in case 1. All units are dimensionless. The energy levels, $E_1=-3.71958$, $E_2=-2.66655$ and $E_4=2.5986$  of the interacting two-boson system (interaction strength  $U_0=-1.0$) are also sketched, with wavy arrows denoting the levels connected by the STIRAP pulses.  Here, $V_0=4.91345043$.}
\label{fig:doublewell}
\end{figure}

\begin{figure} 
\caption{Classical Poincare maps (of  ($p_1,x_1$) for $x_2=1.0$ and $p_2>0$) for two interacting particles in the double well potential. All units are dimensionless. Here $V_0 = 4.91345043$ and $U_0 = -1.0 $. The interaction is approximated by an attractive Gaussian  potential of width $\sigma_c = 0.05$.  (a) Energy $E=E_1=-3.71958$. (b) Energy $E=E_2=-2.66655$. and (c) Energy $E=E_4=2.5986$. A unit area of the phase space equals $\hbar$}
\label{fig:classicalpncr}
\end{figure}

\begin{figure} 
\caption{Classical Poincare maps for lower energies with the bifurcating resonance magnified. All units are dimensionless.A unit area of the phase space equals $\hbar$.(a) Energy $E=E_1=-3.71958$. (b) Energy $E=-3.8$, (c) Energy $E=-4.0$, (d) Energy $E=-4.5$, (e) Energy $E=-4.7$, and (f) Energy $E=-5.0$. Note the increased prominence of the smaller resonance as the energy decreases from $E_1$, as well as the bifurcation in the other resonance as the energy increases from (e).}
\label{fig:bifurcation}
\end{figure}

\begin{figure}
\caption{Plots of energy eigenfunctions for the two interacting bosons in a double well potential. All units are dimensionless. (a1) Contour plot of the  probability density $|\langle x_1,x_2|E_1\rangle|^2$ .  (a2) The cross-section of the wavefunction at $x_2 = -1.0$. (a3) The cross-section of the wavefunction at $x_2 = 0$.  (a4) The cross-section of the wavefunction at $x_2 = +1$. (b1) Contour plot of the  probability density $|\langle x_1,x_2|E_2\rangle|^2$.  (b2) The cross-section of the wavefunction at $x_2 = -1.0$. (b3) The cross-section of the wavefunction at $x_2 = 0$. (b4) The cross-section of the wavefunction at $x_2 = +1$. (c1) Contour plot of the  probability density $|\langle x_1,x_2|E_4\rangle|^2$. (c2) The cross-section of the wavefunction at $x_2 = -1.0$. (c3) The cross-section of the wavefunction at $x_2 = 0$. (c4)  The cross-section of the wavefunction at $x_2 = +1$.}
\label{fig:wavefunctions}
\end{figure}

\begin{figure} 
\caption{Husimi functions  for several energy eigenfunctions for two interacting bosons in a double well potential, with $V_0=4.91345043$ and $U_0=-1.0$. All units are dimensionless. The figures show density plots of the Husimi distribution in the $({x_1} , \bar{p_1})$ plane fixed at $\bar{x_2}=1.0$ and $\bar{p_2}\geq 0$ and subject to classical energy conservation. (a) The energy eigenstate $|E_1{\rangle}$.  (b) The energy eigenstate $|E_2{\rangle}$. (c) The energy eigenstate $|E_4{\rangle}$.}
\label{fig:husimis}
\end{figure}

\begin{figure}
\caption{The STIRAP pulse amplitudes as a function of time for case 1. The first pulse (in time)  connects the intermediate state to the final state of the STIRAP process. All units are dimensionless. The second pulse (in time)   connects the initial state and the intermediate state. The total time $t_{tot}$ is chosen arbitrarily, but the centroids of the pulses are kept at $\frac{t_s}{t_{tot}}=\frac{1}{3}$,$\frac{t_p}{t_{tot}}=\frac{2}{3}$,$\frac{t_{\sigma}}{t_{tot}}=\frac{1}{8}$}
\label{fig:stirap}
\end{figure}

\begin{figure}
\caption{Floquet eigenphase plots for case 1 with $\epsilon_0=10$ and  $\omega=1.05303$. All units are dimensionless.  (a) A  plot of the Floquet eigenphases as a function of $t_{fix}/t_{tot}$. (b) Magnification of the region of interest of in (a).}
\label{fig:e_10phase}
\end{figure}

\begin{figure}
\caption{(a) Plot of $|{\langle}E_j|{\phi}_A{\rangle}|^2$ as a function of $t_{fix}/t_{tot}$ for case 1 with ${\epsilon}_0=10$. All units are dimensionless. The numbers attached to each curve indicate the particular eigenstate $|E_j{\rangle}$ represented. (b) The exact time evolution of $|{\langle}E_j|{\psi}(t){\rangle|}|^2$, obtained by solving the Schr\"odingerer equation with initial state $|\psi(0){\rangle}=|E_1{\rangle}$. The total pulse time is $t_{tot}=24000$.}
\label{fig:e_10states}
\end{figure}

\begin{figure} 
\caption{Floquet eigenphase plots for case 1 with $\epsilon_0=115$ and  $\omega=1.05303$. All units are dimensionless.  (a) A  plot of the Floquet eigenphases as a function of $t_{fix}/t_{tot}$. (b) Magnification of the region of interest of in (a).}
\label{fig:e_115phase}
\end{figure}

\begin{figure} 
\caption{Magnification of avoided crossings in Fig. 12(a). All units are dimensionless. (a) The first avoided crossing is  between the curves for  eigenphases  $\Omega_A$ and $\Omega_C$. (b) A second avoided crossing between $\Omega_A$ and $\Omega_C$ that appears to reverse the effects of the first avoided crossing.}
\label{fig:e_115avcrossing}
\end{figure}

\begin{figure} 
\caption{(a) Plot of $|{\langle}E_j|{\phi}_C{\rangle}|^2$ as a function of $t_{fix}/t_{tot}$ for case 1 with ${\epsilon}_0=115$. All units are dimensionless.  The numbers attached to each curve indicate the particular eigenstate $|E_j{\rangle}$ represented.  (b) Plot of $|{\langle}E_j|{\phi}_A{\rangle}|^2$ as a function of $t_{fix}/t_{tot}$. (c) Plot of $|{\langle}E_j|{\phi}_B{\rangle}|^2$ as a function of $t_{fix}/t_{tot}$.}
\label{fig:e_115states}
\end{figure}

\begin{figure} 
\caption{Plots of $|{\langle}E_j|{\psi}(t){\rangle|}|^2$ as a function of time for case 1 with $\epsilon=115$  and initial condition $|{\psi}(0){\rangle|}=|E_j{\rangle|}$, for different values of  $t_{tot}$. All units are dimensionless. (a) $t_{tot}=72$. (b) $t_{tot}=7000$. (c) $t_{tot}=24000$. (d) $t_{tot}=72000$. (e) $t_{tot}=720000$.}
\label{fig:e_115timeev}
\end{figure}

\clearpage

\begin{figure} 
\caption{Plot of the double-well potential experienced by each boson in case 2. All units are dimensionless. The energy levels, $E_1=-6.42262$, $E_2=-5.68883$ and $E_4=0.640055$  of the interacting two-boson system (interaction strength  $U_0=-1.0$) are also sketched, with wavy arrows denoting the levels connected by the STIRAP pulses. Note the slightly detuned resonance between the $2 \leftrightarrow 4$ and the $4 \leftrightarrow 7$ levels where $E_7=6.96998$. Here, $V_0=7.2912229$.}
\label{fig:doublewell_case02}
\end{figure}

\begin{figure} 
\caption{Classical Poincare maps (of  ($p_1,x_1$) for $x_2=1.0$ and $p_2>0$) for two interacting particles in the double well potential. All units are dimensionless. Here $V_0 = 7.2912229$ and $U_0 = -1.0 $. The interaction is approximated by an attractive Gaussian  potential of width $\sigma_c = 0.005$. Energy $E=E_7=6.96998$. A unit area of the phase space equals $\hbar$.}
\label{fig:classicalpncr_case02}
\end{figure}

\begin{figure} 
\caption{Floquet eigenphase plots for case 2 with $\epsilon_0=115$ and $\omega = 0.091722994$. All units are dimensionless.  The fundamental zone here is $\left\lbrace \frac{-\omega}{2} , \frac{\omega}{2}\right\rbrace $ (a) A  plot of the Floquet eigenphases as a function of $t_{fix}/t_{tot}$. (b) Magnification of the region of interest of in (a).}
\label{fig:phase_case02}
\end{figure}

\begin{figure}
\caption{(a) Plot of $|{\langle}E_j|{\phi}_A{\rangle}|^2$ as a function of $t_{fix}/t_{tot}$ for case 2 with $\epsilon_0 = 115$. All units are dimensionless.  The numbers attached to each curve indicate the particular eigenstate $|E_j{\rangle}$ represented.  (b) Plot of $|{\langle}E_j|{\phi}_B{\rangle}|^2$ as a function of $t_{fix}/t_{tot}$. (c) Plot of $|{\langle}E_j|{\phi}_C{\rangle}|^2$ as a function of $t_{fix}/t_{tot}$.(d) Plot of $|{\langle}E_j|{\phi}_D{\rangle}|^2$ as a function of $t_{fix}/t_{tot}$.}
\label{fig:states_case02}
\end{figure}

\begin{figure} 
\caption{Plots of $|{\langle}E_j|{\psi}(t){\rangle|}|^2$ as a function of time for case 2 with $\epsilon=115$  and initial condition $|{\psi}(0){\rangle|}=|E_j{\rangle|}$, for different values of  $t_{tot}$. All units are dimensionless. (a) $t_{tot}=1000$(b) $t_{tot}=3600$. (c) $t_{tot}=36000$.}
\label{fig:e_115timeev_case02}
\end{figure}

\end{document}